\documentclass{ws-procs9x6}

\newcommand{\be}{\begin{equation}}
\newcommand{\ee}{\end{equation}}

\usepackage{multirow}

\begin{document}

\title{Regge trajectory of the $f_0(500)$ resonance from a dispersive connection to its pole}

\author{J.~Nebreda$^{*,1,2,3,4}$, J.~T.~Londergan$^{2,3}$, J.~R.~Pelaez$^4$ and A.~P.~Szczepaniak$^{2,3,5}$}

\address{$^1$ Yukawa Institute for Theoretical Physics, Kyoto University, Kyoto, 606-8502, Japan\\
$^2$ CEEM, Indiana University, Bloomington, IN 47403, USA\\
$^3$ Physics Department  Indiana University, Bloomington, IN 47405, USA\\
$^4$ Dpto. de F\'isica Te\'orica II, Universidad Complutense de Madrid, 28040, Spain\\
$^5$ Jefferson Laboratory, 12000 Jefferson Avenue, Newport News, VA 23606, USA\\
$^*$E-mail: jnebreda@yukawa.kyoto-u.ac.jp}

\begin{abstract}
We report here our results on how to obtain the Regge trajectory of a resonance from its pole in a
scattering process by imposing analytic constraints in the complex angular momentum plane.
 The method, suited for resonances that dominate an elastic
scattering amplitude, has been applied to the $\rho(770)$ and the $f_0(500)$ resonances.
Whereas for the former we obtain a linear Regge trajectory, characteristic of ordinary
quark-antiquark states, for the latter we find a non-linear trajectory with a much
smaller slope at the resonance mass. This provides a strong indication of the non-ordinary nature of the
sigma meson.
\end{abstract}


\bodymatter

\section{Introduction}\label{aba:sec1}

We make use of the analytical properties of the amplitudes in the complex angular momentum plane to study the Regge trajectories that link resonances of different spins. The form of these trajectories  can be used to discriminate between the underlying QCD mechanisms responsible for generating the resonances. In particular, linear $(J,M^2)$
trajectories relating the angular momentum $J$ and the mass squared are intuitively interpreted in terms of the rotation 
of the flux tube connecting a quark and an antiquark. Strong deviations from this linear behavior would suggest a rather different nature. 

Here we study the trajectory of the lightest resonances in elastic $\pi\pi$ scattering: the $\rho(770)$, which suits well the ordinary meson picture, and the $f_0(500)$ or $\sigma$ meson, whose nature is still controverted and which does not accomodate well in the $(J,M^2)$ trajectories \cite{Anisovich:2000kxa}. 

\section{Regge trajectory from a resonance pole}

An elastic $\pi\pi$ partial wave near a Regge pole reads
\be
t_l(s)  = \beta(s)/(l-\alpha(s)) + f(l,s),
\label{Reggeliket}
\ee
where $f(l,s)$ is a regular function of $l$, and the Regge trajectory $\alpha(s)$ and 
residue $\beta(s)$ are analytic functions, the former having a cut along the real axis for $s > 4m_\pi^2$. 

%

Making use of the analyticity properties of $\alpha(s)$ and $\beta(s)$, and imposing the elastic unitarity condition, we can write down a coupled system of dispersion relations~\cite{Chu:1969ga}:
\begin{align}
\mbox{Re}\, \alpha(s) & =   \alpha_0 + \alpha' s +  \frac{s}{\pi} PV \int_{4m_\pi^2}^\infty ds' \frac{ \mbox{Im}\,\alpha(s')}{s' (s' -s)}, \label{iteration1}\\
\mbox{Im}\,\alpha(s)&=  \frac{ \rho(s)  b_0 \hat s^{\alpha_0 + \alpha' s} }{|\Gamma(\alpha(s) + \frac{3}{2})|}
 \exp\Bigg( - \alpha' s[1-\log(\alpha' s_0)]\nonumber\\
&+ \frac{s}{\pi} PV\!\int_{4m_\pi^2}^\infty\!\!ds' \frac{ \mbox{Im}\,\alpha(s') \log\frac{\hat s}{\hat s'} + \mbox{arg }\Gamma\left(\alpha(s')+\frac{3}{2}\right)}{s' (s' - s)} \Bigg), 
\label{iteration2}
 \end{align}
where $PV$ denotes ``principal value'' and $\alpha_0, \alpha'$ and $b_0$ are free parameters that need to be determined phenomenologically. For the $\sigma$-meson, we modify $\beta(s)$ slightly in order to
include also the Adler-zero required by chiral symmetry. In that case $b_0$ will not be dimensionless.

\section{$\rho(770)$ and $f_0(500)$ trajectories}
  
\begin{figure}
\includegraphics[scale=0.60,angle=-90]{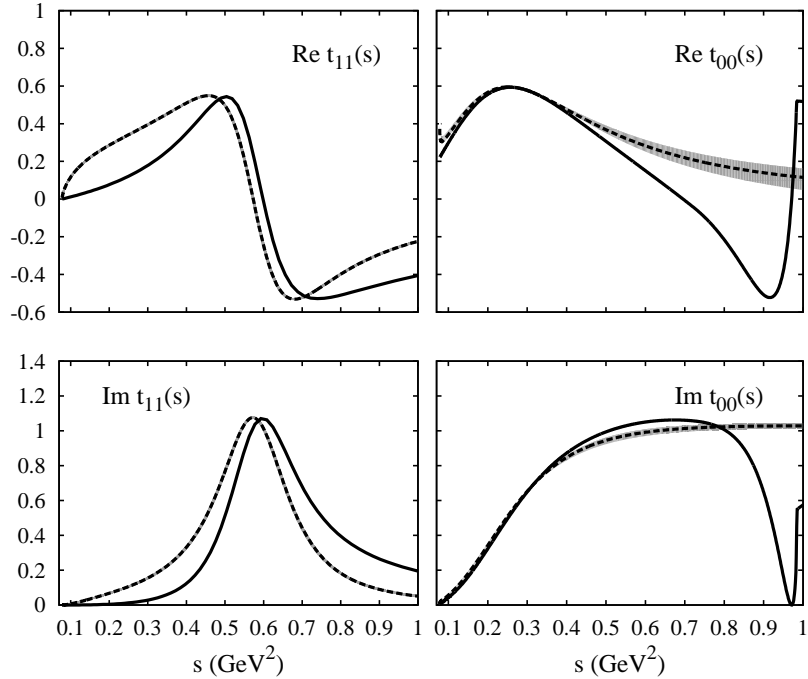}
 \caption{\rm \label{fig:ampl} 
  Partial waves $t_{lI}$ with $l=1$ (left panels) and $l=0$ (right panels). 
  Solid lines represent the amplitudes from \protect{\cite{GarciaMartin:2011jx}}. Their corresponding resonance poles 
  are the input for the constrained Regge-pole amplitudes shown with dashed curves. The gray bands cover the uncertainties due to the errors in the inputs. }
\end{figure}

For a given set of $\alpha_0, \alpha'$ and $b_0$ parameters we solve the system of Eqs.~\eqref{iteration1} and \eqref{iteration2} iteratively. We fix the value of the parameters by a fitting procedure in which we use only three inputs, namely, the real and imaginary parts of the resonance pole position $s_M$ and the absolute value of the residue $|g_M|$. We require that at the pole, on the second Riemann sheet,
$\beta_M(s)/(l  - \alpha_M(s))\rightarrow |g^2_M|/(s-s_M)$, 
with $l=0,1$ for $M=\sigma,\rho$. The pole parameters are taken from a precise dispersive representation of $\pi\pi$ scattering data \cite{GarciaMartin:2011jx}. 
In Fig.~\ref{fig:ampl}  we compare the partial waves that give the input poles\cite{GarciaMartin:2011jx} with the obtained Regge amplitudes on the real axis. They do not need to overlap since they are only constrained to agree at the resonance pole. Nevertheless, we find a fair agreement in the resonant region. As expected, it deteriorates as we approach threshold or the inelastic region, specially in the case of the $S$-wave due to the interference with the $f_0(980)$.


\begin{figure}
\vspace{-3mm}
\hspace{-5mm}\includegraphics[scale=0.55,angle=-90]{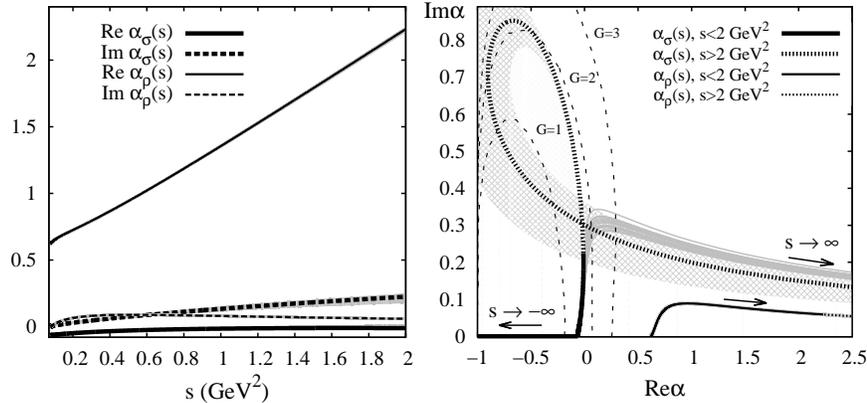}
 \caption{\rm \label{fig:trajectories} 
  (Left) $\alpha_\rho(s)$ and $\alpha_\sigma(s)$ Regge trajectories, 
from our constrained Regge-pole amplitudes. 
 (Right) $\alpha_\sigma(s)$ and $\alpha_\rho(s)$
in the complex plane. 
At low and intermediate energies, the trajectory of the $\sigma$ is similar to those of the Yukawa potential $V(r)=-{\rm G} a \exp(-r/a) /r$ \cite{Lovelace}. For the G=2 curve we can estimate $a\simeq 0.5 \,$GeV$^{-1}$, following \cite{Lovelace}. This could be compared, for instance, to the S-wave $\pi\pi$ scattering length $\simeq 1.6\, $GeV$^{-1}$. }
\end{figure}

We show in the left panel of Fig.~\ref{fig:trajectories} the resulting Regge trajectories, whose parameters are given in Table \ref{aba:tbl2}. The imaginary part of $\alpha_\rho(s)$ is much smaller than the real part, and the latter grows linearly with $s$. Taking into account our approximations, and that our error bands only reflect
the uncertainty in the input pole parameters, the agreement with previous determinations is remarkable: $\alpha_\rho(0)=0.52\pm0.02$~\cite{Pelaez:2003ky}, $\alpha_\rho(0)=0.450\pm0.005$ 
\cite{PDG}, $\alpha'_\rho\simeq 0.83\,$GeV$^{-2}$ \cite{Anisovich:2000kxa}, $\alpha'_\rho=0.9\,$GeV$^{-2}$ \cite{Pelaez:2003ky}, or $\alpha'_\rho\simeq 0.87\pm0.06$GeV$^{-2}$, \cite{Masjuan:2012gc}.

\begin{table}
\vspace{-3mm}
\begin{center}
\tbl{Parameters of the $\rho(770)$ and $f_0(500)$ trajectories}
{\begin{tabular}{cccc}\toprule
& $\alpha_0$ & $\alpha'$ (GeV$^{-2}$)  & \hspace{5mm}$b_0$\hspace{5mm} \\\colrule
$\rho(770)$ & $0.520\pm0.002$  & $0.902\pm0.004$ & $0.52$ \\
$f_0(500)$ &  $-0.090\,^{+\,0.004}_{-\,0.012}$ & $0.002^{+0.050}_{-0.001}$ & $0.12$ GeV$^{-2}$\\
\botrule
\end{tabular}}
\label{aba:tbl2}
\end{center}
\vspace{-4mm}
\end{table}

However, the $f_0(500)$ trajectory is evidently nonlinear and its slope is about two orders of magnitude smaller than that of the typical to quark-antiquark  resonances, {\it e.g.}, $\rho$, $a_2$, $f_2$ or $\pi_2$. 
 This provides strong support for a non-ordinary nature of the $\sigma$ meson. Moreover, the  
 tiny slope excludes the possibility 
 that any of the known isoscalar resonances lie on its  trajectory.

Furthermore, in Fig.~\ref{fig:trajectories} we show the striking similarities
 between the $f_0(500)$ trajectory and those of Yukawa potentials
in non-relativistic scattering. Of course, our results are most reliable at low energies (thick dashed-dotted line) and the extrapolation should be interpreted cautiously. Nevertheless, our results suggest that the $f_0(500)$ looks more like a low-energy resonance of a short range potential,  {\it e.g.}\ between pions,  than a bound state of a long range confining force between a quark and an antiquark.

\bibliographystyle{ws-procs9x6}

\end{document}